\documentclass[10pt,a4paper]{article}
\usepackage{subfigure}
\usepackage{graphicx}
\usepackage{subfloat}
\usepackage{lineno}
\usepackage{comment}
\usepackage{amssymb,amsmath}

\usepackage[sc]{mathpazo} 
\usepackage[T1]{fontenc} 
\usepackage{microtype} 
\usepackage{hyperref} 

\usepackage[hmarginratio=1:1,top=32mm,columnsep=20pt]{geometry} 
\usepackage[hang, small,labelfont=bf,up,textfont=it,up]{caption} 

\usepackage{abstract} 

\usepackage{titlesec} 
\titleformat{\section}[block]{\large\scshape\centering{\Roman{section}.}}{}{1em}{} 

\usepackage{fancyhdr} 
\begin{document}

\title{\vspace{-15mm}\fontsize{14pt}{10pt}\selectfont\textbf{Measurements of Cosmic Rays with IceTop/IceCube:\\Status and Results}} 

\author{
\large
\textsc{Alessio Tamburro}\\
\footnotesize Bartol Research Institute and Department of Physics and Astronomy,\\
\footnotesize University of Delaware, Newark, DE 19716, USA\\
\footnotesize \href{mailto:atamburro@icecube.wisc.edu}{atamburro@icecube.wisc.edu} 
\vspace{-5mm}
}
\date{}

\maketitle 

\thispagestyle{fancy} 


\begin{abstract}

\noindent 
The IceCube Observatory at the South Pole is composed of a cubic kilometer scale neutrino telescope buried beneath the icecap and a square-kilometer surface water Cherenkov tank detector array known as IceTop.
The combination of the surface array with the in-ice detector allows the dominantly electromagnetic signal of air showers at the surface and their high-energy muon signal in the ice to be measured in coincidence. This ratio is known to carry information about the nuclear composition of the primary cosmic rays. This paper reviews 
the recent results from cosmic-ray measurements performed with  IceTop/IceCube: energy spectrum, mass composition, anisotropy, search for PeV $\gamma$ sources, detection of high energy muons to probe the initial stages of the air shower development, and study of transient events using IceTop in scaler mode.

\end{abstract}



\section{Introduction}
\label{sec::intro}

For 100 years since their discovery, primary cosmic rays 
have been measured with different techniques
and at different locations on Earth\cite{Kampert:2012vi}.
Particularly interesting are those cosmic particles 
in the energy range between 3~$\cdot 10^{14}$~eV (300~TeV or 0.3~PeV) 
and $10^{21}$~eV (1000~EeV), whose
origin, composition, and energy spectrum 
remain not fully understood. Cosmic magnetic fields permeating all
space prevent the localization
of the sources that produced the charged particles observed at Earth.
In addition, the relatively low flux of these primaries (about 1~m$^{-2}\cdot$yr$^{-1}$ at $10^{15}$~eV)
does not allow for direct measurements.
Subsequent interactions in the atmosphere result in air showers of secondary
particles that are sampled at ground. These secondaries also release 
Cherenkov and fluorescence light detected with telescopes.

\begin{figure}[t]
\begin{center}
\includegraphics[width=10.0cm]{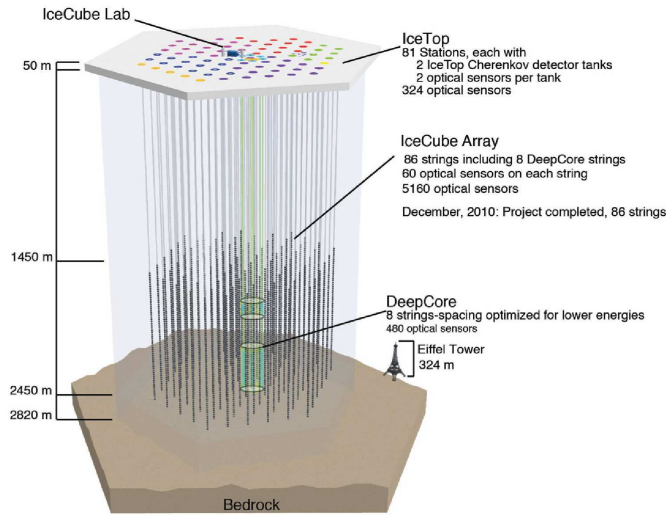}
\caption{Sketch of the IceCube Observatory labeled with some of its main features.
The different colors at the surface identify different deployment stages of 
the detector. IceCube in its 2006-07 configuration is shown in red
and termed as IT26/IC22 (26 IceTop stations/22 in-ice strings). 
Other configurations are IT40/IC40 (2007-08) in green, 
IT59/IC59 (2008-09) in violet, IT73/IC79 (2009-10) in blue, and
IT81/IC86 (2010-11) in yellow.
\protect\label{fig::icecube}}
\end{center}
\end{figure}

The IceCube Observatory is a three-dimensional cosmic-ray air shower
detector (Fig.~\ref{fig::icecube}). The surface component, IceTop, is
an array of water Cherenkov tanks that samples the shower at the surface.
The deep detectors of IceCube (between 1.45~km and 2.45~km below
the surface) measure the signal from penetrating muons, which have 
500~GeV or greater at production in the atmosphere.
Events seen in coincidence by both the surface and the deep detectors (coincident events)
offer a unique view of cosmic-ray air showers because
they carry information from  two different regions of shower development.
The surface shower is
dominated by low-energy photons, electrons, and muons produced throughout
the cascade, while the high-energy muons reflect the early stage of shower development.
In addition to coincident events, it is also possible to use IceTop as a stand-alone air shower array and to
use the deep detector by itself as a muon detector. Together, the three types of events allow
measurement of the energy spectrum, composition, and anisotropy of the primary cosmic rays
from about 10~TeV to about 1~EeV.

The main goal of IceCube is to detect and measure high-energy neutrinos of extra-terrestrial origin,
as described in the companion review article in this journal\cite{Taboada:2012}.
The high-energy neutrinos are expected to point back to the cosmic-ray sources 
in which they are produced\cite{Gaisser:2012ru}.
From the point of view of neutrino astronomy, IceTop serves as a partial {\it veto} against
atmospheric background in the deep detector.

The emphasis of this paper is to review the apparatus and its performances, and the recent physics results
of IceCube as a cosmic-ray detector.
The performances will be reviewed in Sec.~\ref{sec::det}. 
Current results from cosmic-ray mass composition and energy spectrum
analysis will be discussed in Sec.~\ref{sec::comp}.
The anisotropy measured with IceTop data is presented in Sec.~\ref{sec::ani}. 
Finally, this paper will also report on recent results including $PeV$ $\gamma$ 
search (Sec.~\ref{sec::pevs}), high momentum in-ice laterally separated muons
(Sec.~\ref{sec::lsm}), and detection of transient events (solar flares and gamma-ray bursts)
with IceTop (Sec.~\ref{sec::moni}).

\section{Cosmic-ray Detection and Reconstruction}
\label{sec::det}

The IceTop detector\cite{IT} is a surface array of 162 cylindrical Cherenkov tanks 
installed at the 2835~m altitude of the South Pole surface (atmospheric depth 
of about 680~g/cm$^2$). The tanks are 1.86~m in diameter and 1.10~m in height.
Filled with clear ice to a depth of 0.90~m, they operate on the same principle as the water tanks
of the Haverah Park experiment\cite{HP} and the Pierre Auger Observatory\cite{PA}.
To minimize accumulation of drifting snow, the tanks have their top surface 
level with the surrounding snow. Nevertheless, a variable
and not negligible snow coverage is measured and accounted for when analyzing data.

Pairs of tanks, 10~m apart from each other, localize 81 stations that
are distributed over an area of about 1~km$^2$ 
on a triangular grid with mean spacing of approximately 125~m.
An {\it in-fill} array for denser shower sampling is configured in the center of IceTop
by using 3 stations at smaller distances together with their 5 neighboring stations.
Each IceTop tank contains two standard IceCube
digital optical modules\cite{DOM,fyp} (DOMs), which include light sensor and readout electronics.
To enhance the dynamic range, the DOMs
of each tank are run at 2 different gains (low and high) with resulting effective thresholds
of 20 and 200 photoelectrons (pe), respectively.
Cosmic-ray muons, hitting the tanks at an approximate rate of 2~kHz,
provide the basic calibration of tanks. The signal spectrum of a tank consists of a
low-energy electromagnetic component and a muon peak at higher energy.
The tank signals are calibrated in units of ``vertical equivalent muon'' or VEM
whose definition is based on the muon peak of the tank spectrum.
A vertically through-going muon of about 1~GeV produces approximately 125~pe
in a high-gain DOM. 
In addition to measuring signal amplitudes, the IceTop DOMs can record
the counting rate of low-energy cosmic rays (scaler mode). The rates are available
for heliospheric studies of solar modulation and transient events 
such as solar flares and gamma-ray bursts (GRBs).

To initiate the readout of a DOM, its neighbor in the other tank
at the same station has to be hit (hard local coincidence or HLC).
This suppresses the background of accidental signals
caused by isolated muons and allows a good angular resolution.
The basic IceTop air shower trigger requires 6 DOMs to 
``launch'' or report signals within a 5~$\mu$s time window.
Thus, even if only high-gain DOMs are
above threshold, this trigger includes all 3-station events.
The rate for these events is 30~Hz.
Once the trigger is formed, additional signals from single tanks 
are added if they occur within a 20~$\mu$s time window
(soft local coincidence or SLC). Counting SLCs is useful to evaluate the shower muon content.
These SLC signals are also read out when an in-ice
trigger forms either with or without the presence of HLC signals. 
In this case, SLCs can be used to veto
dow-going cosmic rays when measuring up-going neutrinos.

\begin{figure}
  \begin{minipage}[]{0.42\textwidth}
    \centering
   \subfigure[]{\includegraphics[width=\textwidth]{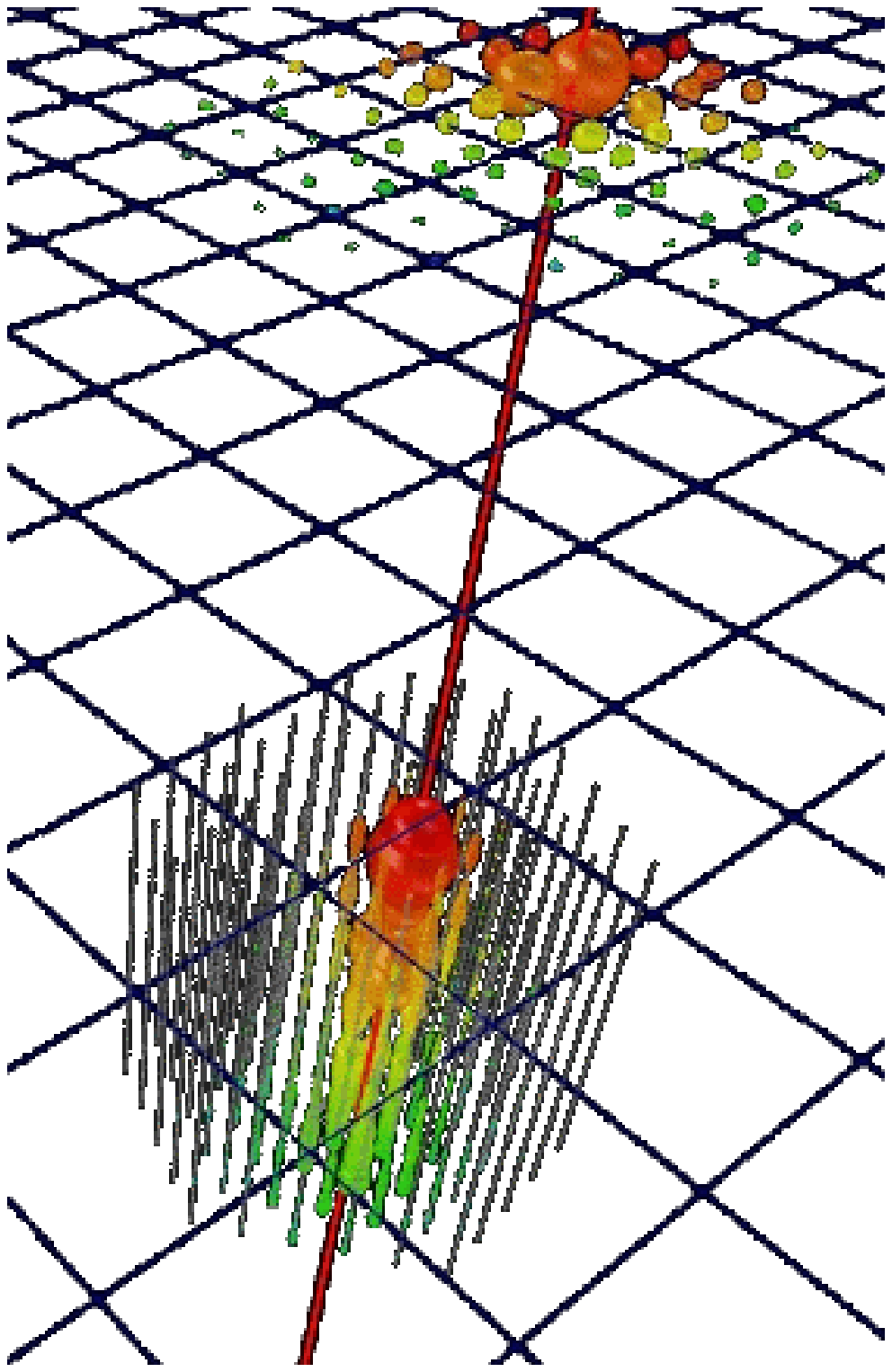}\protect\label{fig::coinceve}}
 \end{minipage}\hfill
 \begin{minipage}[]{0.46\textwidth}
   \centering
   \subfigure[]{\includegraphics[width=0.88\textwidth]{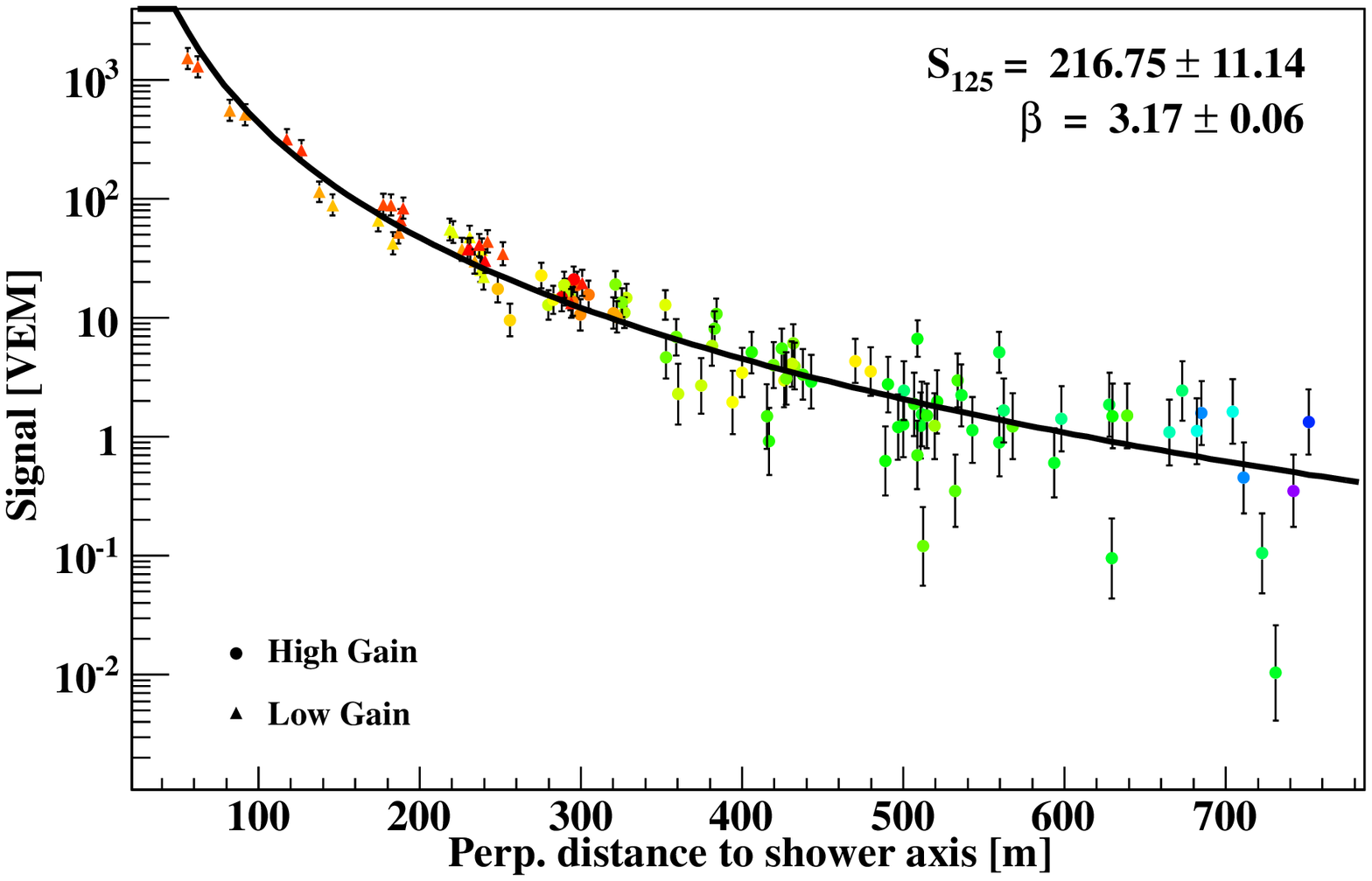}\protect\label{fig::dlp}}\\
   \subfigure[]{\includegraphics[width=0.85\textwidth]{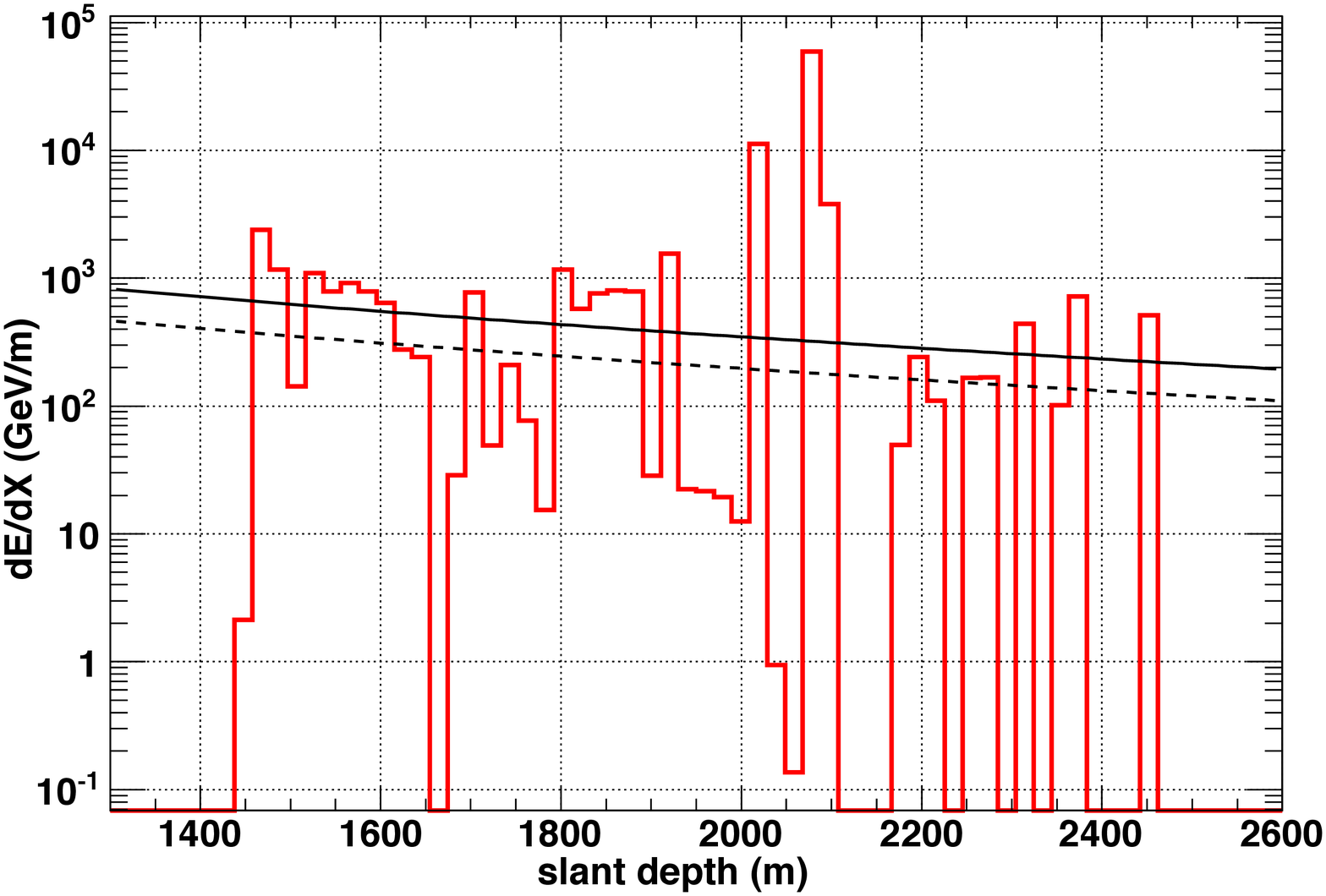}\protect\label{fig::dedx}}
 \end{minipage}\hfill
 \caption{\label{Fig: anis}{(a) Coincident event recorded by IceTop/IceCube in 2010 (IT73/IC79). The triggered
 DOMs are indicated with colored spheres with radii proportional to the signal. (b) Lateral distribution of tank
 signals in VEM for the event in Fig.~\ref{fig::coinceve} fitted to a double logarithmic parabola
 (see text for details). (c) Muon energy loss $dE/dX$ as a function of the in-ice depth 
 for the event in Fig.~\ref{fig::coinceve}. The stochastic energy loss is about 880~GeV/m above the fitted average energy
 loss (black line). The dashed line is the average energy
 loss after removal of the stochastic peaks.}}
\end{figure}

There are different topologies of events that are
relevant for analysis of cosmic-ray data
with IceTop/IceCube. This paper will concentrate on results obtained with analysis of
events caused by nearly vertical (zenith angles $\theta\lesssim$~37$^\circ$ or $\cos\theta >$~0.8) 
and contained air showers (Fig.~\ref{fig::coinceve}).
These showers are reconstructed with their axis crossing both parts of the detector. 
The effective area of IceCube for such coincident events is
A$\approx$0.15~km$^2$sr. 
The maximum energy above which the intensity is too low to obtain
enough events for analysis is about 1~EeV. 
The effective area rises to 0.4~km$^2$sr 
for detecting events by IceTop alone.

The surface shower particle density decreases rapidly with the distance from
the shower axis (lateral distribution function). 
This lateral distribution carries
information about the energy of the primary particles.
The charge expectation value $S$ in an IceTop tank at distance $r$ from the shower axis is described 
with a ``double logarithmic parabola'' (Fig.~\ref{fig::dlp}) as follows\cite{Klepser}
\begin{equation}
\label{eq::dlp}
S(r) = S_{ref}\cdot \left(\frac{r}{R_{ref}}\right)^{-\beta-\kappa\log_{10}(r/R_{ref})},
\end{equation}
where $R_{ref}=~$125~m, $S_{ref}$ is the charge in VEM at $R_{ref}$, and $\kappa=$~0.303.
The parameter $S_{ref}$ is thus referred to as $S_{125}$ and is a measurement
of the shower size. The signals measured
between about 30~m and 300~m from the shower axis are quite well
described by Eq.~\ref{eq::dlp} for primary
energies in the range 1--100~PeV and arrival directions with zenith in the range
0$^\circ$--40$^\circ$. 

The arrival time behind the shower plane, $\Delta t(r)$, as a function
of the lateral distance $r$ from the shower axis is described
by the sum of a parabola and a Gaussian function, both symmetric around the
shower axis, with the following form 
\begin{equation}
\label{eq::artimes}
\Delta t(r) = ar^2 + b\left( \exp\left( \frac{r^2}{2\sigma^2_r}\right)-1\right),
\end{equation}
where $a =$~4.823$\cdot$10$^{-4}$~ns/m$^2$, $b =$~19.41~ns, and $\sigma_{r}=$~83.5~m.
The energy, zenith angle, and mass dependence of $a$, $b$, and $\sigma_r$ is currently
under investigation in simulations.
When reconstructing the shower direction $\vec{n}$, Eq.~\ref{eq::artimes} 
accounts for the shower front curvature as follows
\begin{equation}
\label{eq::shplane}
t(\vec{x}) = t_0 + \frac{\vec{x} - \vec{x}_c}{c} \vec{n}+ \Delta t (r),
\end{equation}
where $t(\vec{x})$ is the tank signal time, 
$\vec{x}$ the tank position, $\vec{x}_c$ the shower core position,
and $t_0$ the time when the core reaches the surface. 
For events triggering 5 or more stations, Eq.~\ref{eq::shplane} is fitted to
the measured signal times with 5 free parameters, 3 for core position and time, 
and 2 for shower direction. 
At the same time, Eq.~\ref{eq::dlp} is fitted to the measured tank signals with
2 free parameters, $S_{125}$ and $\beta$. A maximum likelihood method is 
adopted to obtain the best fit to the measurements\cite{IT}. 
For events triggering 3 or 4
stations, only $S_{125}$ and $\beta$ are kept free after fitting the measured signal times
to obtain core position and time, and shower direction.

The position of IceTop at the high altitude of the South Pole
makes it possible to sample ground particles near the shower maximum,
thus reducing significantly the effects of fluctuations
and allowing accurate measurements.
Events of 5 or more stations are expected from primaries of
energy between PeV and EeV, whereas 3 or 4 station
events set the threshold for cosmic ray detection with IceTop to 
about 300~TeV. Lower energies ($\gtrsim$100~TeV) are expected if the events triggering 
the denser in-fill array are considered.
The energy resolution is estimated to be $\lesssim$0.1 in units of $\log_{10}($E/GeV$)$
above about 1~PeV, reaching about 0.05 for $E>$~10~PeV.
The angular resolution ranges from about 1$^\circ$ at 10$^{15}$~eV to 
about 0.2$^\circ$ at 10$^{17}$~eV. The core resolution ranges from about 15.0~m at 10$^{15}$~eV
to about 6~m at 10$^{17}$~eV.

Analysis of IceCube coincident events aims to give clear insights 
into the nuclear composition of cosmic rays
for energies that span from PeV to EeV.
For a given primary energy, penetrating muons are more abundant
in iron showers than in proton showers since their development starts higher in the atmosphere.
The muon bundle size is therefore larger for iron showers.
On the other hand, proton showers reach their maximum development
deeper in the atmosphere than iron showers and the ratio of muons to electrons and
photons is therefore smaller at the surface.  
The number of in-ice muons or muon multiplicity is closely related to the
amount of energy deposited in the detector, which is proportional to the
amount of Cherenkov light generated. 
For example, depending on the mass of the primary particle (proton or iron),
an event of  5$\cdot$10$^{15}$~eV is expected to carry 30 to 80 muons with
sufficient energy to reach  a depth of 1500~m and deposit 5$\cdot$10$^{12}$~eV to 15$\cdot$10$^{12}$~eV in the deep detector.  
Analogous to surface measurements of tank signals,
the in-ice DOM signals (in photoelectrons)
can be described in terms of a lateral distribution, which
is a function of the distance from the muon bundle track
at a given slant depth from the surface.
This function is dominated by a decaying exponential
whose slope is the attenuation length of light in the ice\cite{IceCube:2012vv}.
The bundle size, termed as $K_{70}$, is defined as the in-ice signal measured
at a slant depth of 1950~m and at a perpendicular distance from
the track of 70~m. This observable has been used in combination
with $S_{125}$ to discriminate between light and heavy primary masses (see Sec.~\ref{sec::comp}).
A different approach explored the reconstruction of the muon bundle
energy loss $dE/dx$ to find composition sensitive properties
of bundles\cite{Feusels:2009xb}. The muon bundle energy loss as a function of the slant depth
is a convolution of the shower muon multiplicity, 
the muon energy distribution and the energy
loss of a single muon. For iron showers,
the bundle energy loss is expected to be greater.
Furthermore, for the same amount of energy
deposited, stochastic losses along the 
bundle track are larger in proton showers, which 
produce more high energy muons.

The reconstruction of air showers with IceTop/IceCube is affected by
several uncertainties. These uncertainties have been extensively investigated
and several advances are expected in the future results.
Pressure variations at the surface and snow coverage affect 
the measurement of $S_{125}$ with the latter giving the largest contribution.
For a given primary energy and arrival direction, higher pressure
makes the primary interaction depth shallower thus reducing
the shower size. 
The snow over each tank is physically measured every year 
and also estimated from the muon/electron ratio in calibration curves. 
The snow mainly affects the response of the tank to
the electromagnetic component of the shower front thus affecting
the energy threshold. 
The shower size is corrected
for pressure and snow coverage.
Other systematic uncertainties affect the in-ice reconstruction: the ice model used to describe the 
properties of the photon propagation\cite{SpiceMie} ($\pm$10\%), 
the DOM efficiency\cite{Abbasi:2010vc} ($\pm$10\%),
and the muon rate seasonal variation\cite{Tilav:2010hj}.

\section{Cosmic-ray Energy Spectrum and Nuclear Composition}
\label{sec::comp}

Most galactic cosmic rays are believed to be accelerated 
in the blast waves (diffusive shocks) of nearby supernova remnants\cite{Hillas:2005cs} (SNRs)
and in some cases (extended sources/strong magnetic fields) 
can escape the acceleration region with energies
up to about 10$^{18}$~eV. 
The signature of these sources is the gradual steepening
of the cosmic-ray flux at a few 10$^{15}$~eV, called {\it knee}.
The well-known power-law form of the
cosmic-ray energy spectrum $dN/dE\propto E^{-2.7}$ changes
its spectral index to about -3.0.
The knee is interpreted as the maximum energy reached through
acceleration in SNRs. This energy scales with the charge of the nucleus.
Therefore, the spectrum of galactic cosmic rays
is predicted to end at energies of about 10$^{18}$~eV with
the heaviest elements being accelerated\cite{Blasi:2011fi}.
Depending on whether the transition is from galactic iron to extragalactic
proton or from galactic iron to extra-galactic mixed composition of different nuclei,
the transition energy is predicted\cite{Berezinsky:2007wf} at  few 10$^{17}$~eV or 
few 10$^{18}$~eV. 
At energies between 10$^{15}$~eV to 10$^{17}$~eV, 
all air shower experiments observe energy dependent changes of
composition that are compatible with an increase of the measured average mass
of cosmic rays. Above 10$^{17}$~eV and up to 10$^{18}$~eV
measurements of composition indicate a decrease 
of the cosmic-ray average mass\cite{Kampert:2012mx}.
The cosmic-ray flux up to 10$^{17}$~eV is believed to be mostly dominated by
the contribution of galactic sources.
The ``fingerprint'' of the transition to the extra-galactic contribution 
is expected in the measurement of the
mass composition at this energy and above.
At the highest energies (above 10$^{19}$~eV), measurements from
the Pierre Auger Observatory, and HiRes/Telescope Array 
give opposite results\cite{Abraham:2010yv,Abbasi:2009nf,Sagawa:2011zza}.
These experiments are extending their detectors to 
reach about 10$^{18}$~eV and below\cite{Maris:2011zz,Mathes:2011zz,Ogio:2011zz},
and their measurements will overlap with IceCube measurements.

IceCube/IceTop allows for precise measurements of the primary energy spectrum
in a wide energy range that reaches the energy threshold of 
the largest air shower arrays and is sensitive to cosmic-ray nuclear composition changes.
\begin{figure}[t]
\centering
\includegraphics[width=10.0cm]{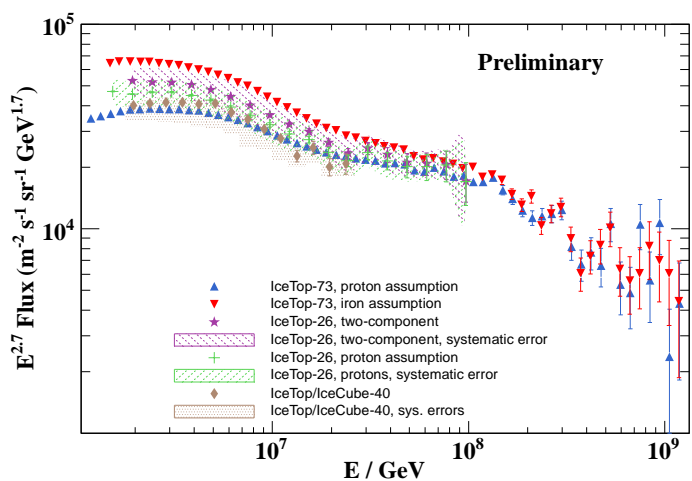}
\caption{Energy spectra obtained with data of IceTop running in different configurations (IT26, 
IT40, and IT73). For the IT26 and IT40 spectrum, the systematic uncertainty is also shown.
\protect\label{fig::spectraIT}}
\end{figure}
The first analysis to determine the all-particle energy spectrum
with IceTop\cite{Abbasi:2012wn}
was based on data of IT26 (area of 0.094~km$^2$) 
taken between June and October 2007. 
In this analysis, the measured shower size ($S_{125}$) spectra in three zenith angle 
ranges up to 46$^\circ$ are unfolded or de-convoluted
into the estimated energy spectrum.
The spectrum obtained in the energy range 
between 1 and 100~PeV is shown in Fig.~\ref{fig::spectraIT} along with other spectra measured 
with IceTop data and discussed in this section.
The energy at which the lines with the minimum and maximum slope before and after the
knee intersect identifies the knee position.   
Assuming pure iron, the energy spectra measured in different zenith angle ranges
have been shown to disagree and prove that pure iron primaries can
be excluded in the energy range up to 25~PeV.
A consistent interpretation of the spectra
measured at different zenith angles requires a mixed composition. 
For a two-component model, 
the knee measured in the IT26 spectrum is at 4.32~PeV and
the spectral index above the knee is -3.11. An indication of a flattening of
the spectrum above 22~PeV is also observed with a spectral index changing to -2.85.

A preliminary measurement of the spectrum with IT73 has been 
obtained by analyzing 11 months of data (June 15 to 
May 13, 2010)\cite{Tamburro:2012,Serap:2012,Bakhtiyar:2012} (Fig.~\ref{fig::spectraIT}).
The statistics is nearly 40,000,000 events between about 
0.3~PeV and 1~EeV and for $\cos\theta>$~0.8. Of these events,
about 200 are found above about 200~PeV.
With enhanced precision, better energy resolution, and larger statistics above 100~PeV, 
this measurement confirms the earlier result of a flattening observed with IT26 data
and reveals the spectrum structure in the steepening between the knee and ankle 
above about 100~PeV. It also appears that the spectrum is not well described by a single 
power law. All showers with 5 or more stations and with $\cos\theta >$~0.8 are considered in this analysis.
The energy spectrum, now measured between 1~PeV and 1~EeV, will
be extended to lower energies, down to 300~TeV.
This can be achieved by selecting small events in IceTop\cite{Bakhtiyar:2009}
and is particularly interesting in view of the recent
measurements with ATIC\cite{Panov:2011ak}
and CREAM\cite{Yoon:2011aa} balloon-borne calorimeters that are providing increased
statistics and a new view of the region around 100~TeV/nucleus.

\begin{figure}[t]
\centering
\includegraphics[width=10.0cm]{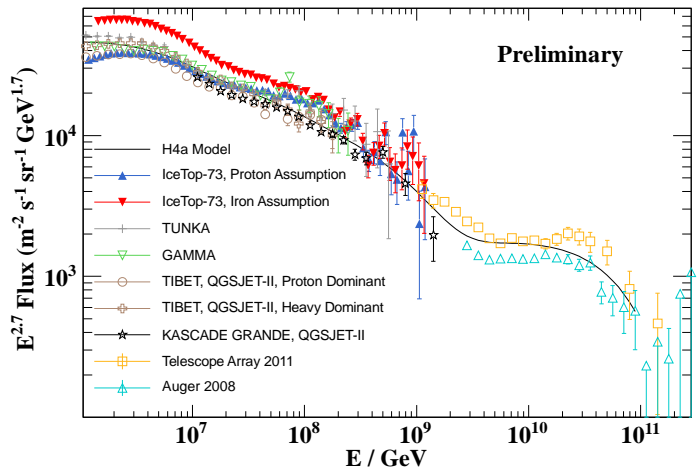}
\caption{Cosmic-ray energy spectrum measured with IT73 and other experiments operating 
in the same energy range. The spectra from the Pierre Auger Observatory\protect\cite{Augerspec}
and Telescope Array\protect\cite{TAspec} are also shown. Finally, the {\it H4a} model of the cosmic-ray flux 
is overlaid (see text for detail).}
\protect\label{fig::spectraAll}
\end{figure}
In Fig.~\ref{fig::spectraAll}, the IceTop spectrum is compared to the spectra obtained
with recent measurements of
KASCADE-Grande (Karlsruhe, Germany, 110~m  a.s.l.)\cite{Apel:2011mi,Apel:2012rm}, 
Tibet Array (at Tibet Yangbajing, 4300~m  a.s.l.)\cite{Amenomori:2008aa,Amenomori:2011zza}, 
GAMMA (on the south side of Mount Aragats in Armenia, 3200~m a.s.l.)\cite{Garyaka:2008gs}, 
and Tunka (in the Tunka Valley in Buryatia, Siberia, 675~m a.s.l.)\cite{Tunka:spectrum}.
A model assuming three populations of cosmic rays\cite{Gaisser:2012zz} 
(SNR component,  high energy
galactic component, and extra-galactic component) 
and termed as {\it H4a} is also shown.

\begin{figure}[t]
\centering
\includegraphics[width=10.0cm]{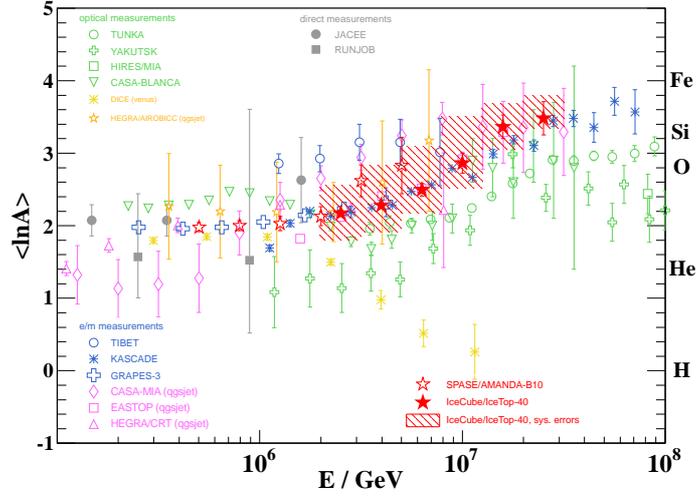}
\caption{The mean logarithmic mass vs primary energy adapted from 
Ref.~\cite{IceCube:2012vv}. 
The IT40/IC40 results are shown with (red) stars along with
their statistical errors (solid red error bars) and systematic errors (shaded red region).
Measurements from other experiments are also shown. Data points are compiled from
Ref.~\cite{Hoerandel:2002yg,Kampert:2012mx}.}
\protect\label{fig::massc}
\end{figure}

A first attempt to measure the mass composition of cosmic rays
was performed with one month of data (constant snow coverage) of IceCube in its 2008 IT40/IC40
configuration\cite{IceCube:2012vv}.
A neural network was trained with Monte Carlo simulations
of 5 primaries (proton, helium, oxygen, silicon, and iron).
Measurements of the electromagnetic
component of the air showers at the surface ($S_{125}$) and the muon component 
in the ice ($K_{70}$) are used to ``teach'' the network how
to find the best fit to the primary energy and mass. 
A measurement of the cosmic-ray energy spectrum (Fig.~\ref{fig::spectraIT}) and
composition (Fig.~\ref{fig::massc}) at energies between 1~PeV and 30~PeV was 
determined. 
The energy resolution from the neural network ranges from 18--20\% in the threshold region
of this analysis (1.5~PeV) to 6--8\% at 30~PeV, 
for an average resolution better than 14\% over the full range of
energies. 
The energy spectrum derived with this analysis is consistent with other
IceTop results within systematic uncertainties.
The mean logarithmic mass $<lnA>$ as a function of the primary energy 
indicates a strong increase in mass through the knee
although the systematic uncertainties can greatly affect the measured composition in terms
of absolute value of $<lnA>$.

\section{Cosmic-ray Anisotropy Measurements}
\label{sec::ani}

Although detection of $\gamma$ rays and neutrinos from individual galactic
or extra-galactic sources of cosmic rays remains a method to  
probe the origin of cosmic rays, studies of anisotropy of cosmic-ray arrival directions
are important to investigate the characteristics of cosmic-ray  
propagation in the local interstellar medium. 

In IceCube the anisotropy of cosmic-ray arrival directions can be measured
in two different ways: using TeV muon events collected with the deep detector
or using cosmic-ray air shower events triggering the surface array.
The in-ice detector has a lower energy threshold than IceTop, which allows  
to investigate the anisotropy of cosmic rays at lower primary energies (down to 
20~TeV) and larger zenith angles (up to 90$^\circ$).
This makes it possible for the deep detector to
reach a higher sensitivity (about 6.3$\cdot$10$^{10}$~events/yr, 
anisotropy level $\delta >$~10$^{-5}$) and scan small scale structures.
On the other hand, IceTop presents a better energy resolution (20\% at $>$300~TeV),
although binning is limited by statistics, and potential for including
composition sensitivity. 

A dipole-like large scale anisotropy (amplitude of about 10$^{-3}$) 
from few tens of TeV to about 100~TeV
has been observed with in-ice data\cite{Abbasi:2010mf,Abbasi:2011zka}.
This anisotropy is inconsistent, both in 
amplitude and phase, with the Compton-Getting prediction\cite{CG},
i.e. the apparent anisotropy caused by the relative motion between the Earth and sources
of cosmic rays. A small scale anisotropy with
significant structure at angular sizes between 10$^\circ$
and 30$^\circ$ has been also observed.
This might uncover non-diffusive propagation effects or SNR connection
and be a natural consequence of the stochastic nature of cosmic-ray galactic sources, in
particular nearby and recent SNRs ($<$0.1--1~kpc)\cite{Erlykin:2006ri,Blasi:2011fl,Biermann:2012tc}.

\begin{figure}[t]
  \begin{minipage}[]{0.48\textwidth}
     \centering
   \subfigure[]{\includegraphics[width=\textwidth]{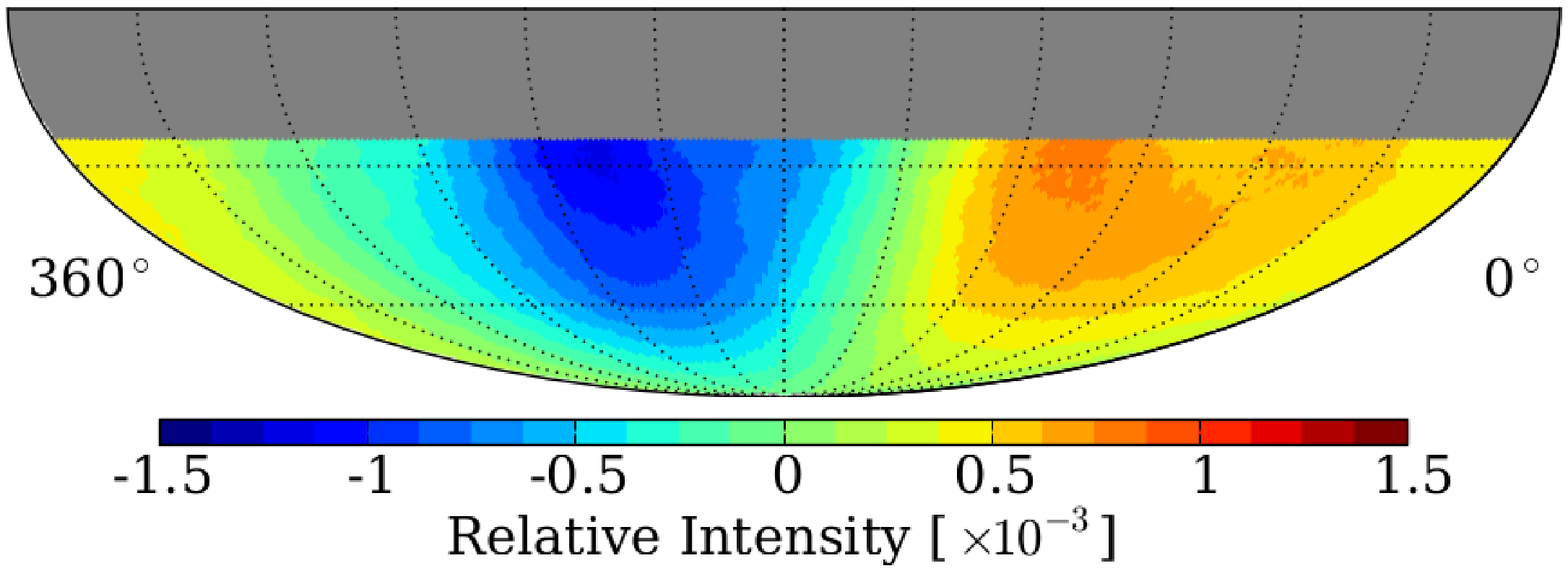}}
    \subfigure[]{\includegraphics[width=\textwidth]{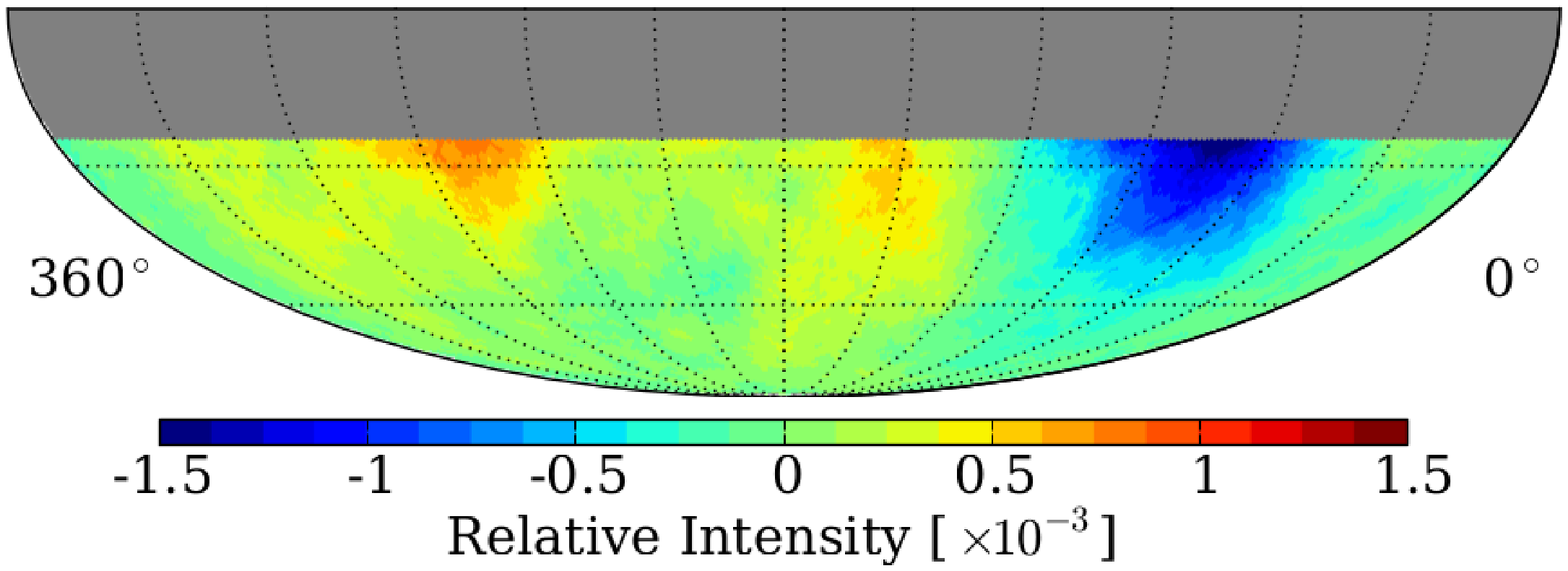}}
 \end{minipage}
\hfill
 \begin{minipage}[]{0.48\textwidth}
    \centering
\subfigure[]{\includegraphics[width=\textwidth]{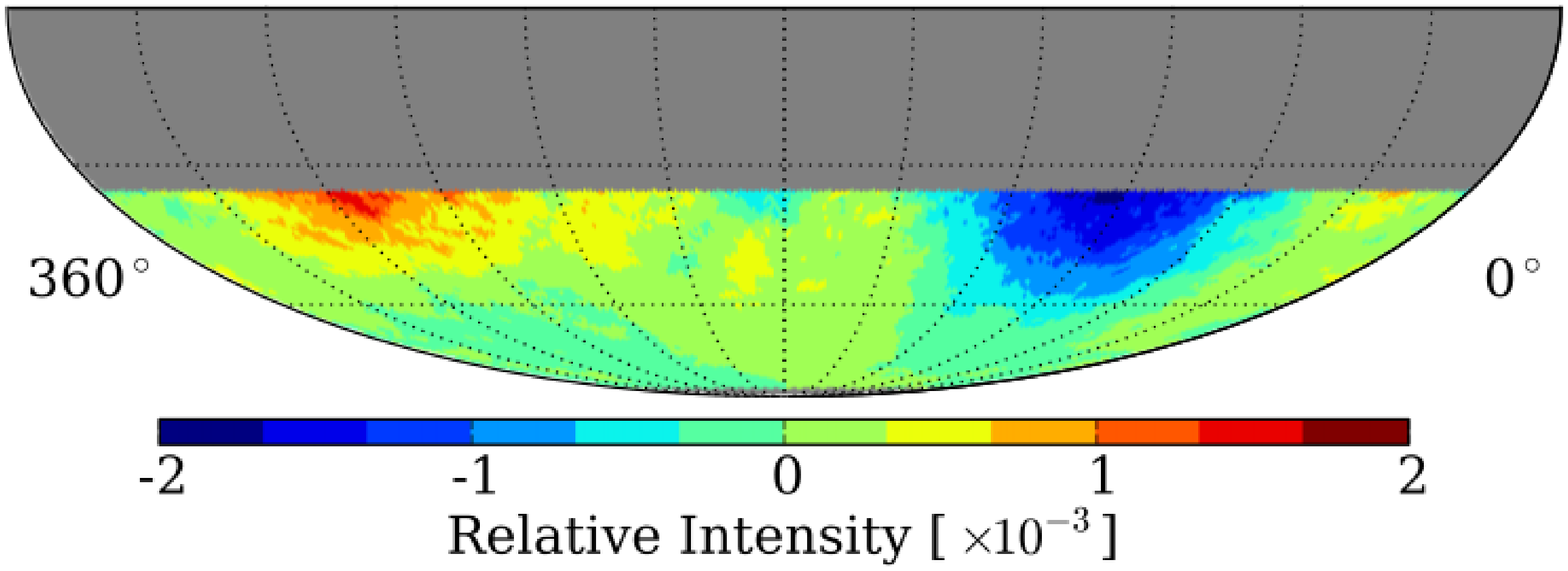}}
\subfigure[]{\includegraphics[width=\textwidth]{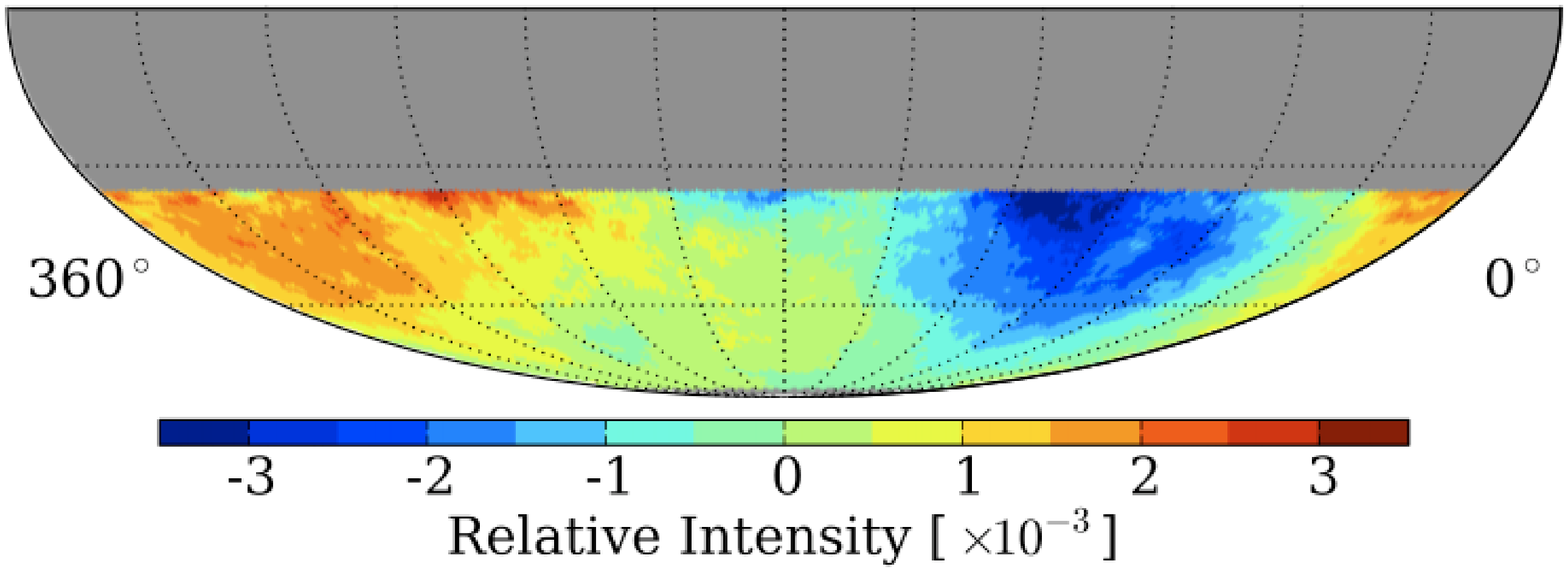}}
 \end{minipage}\hfill
 \caption{\protect\label{fig::skymaps}{Statistical significance sky maps obtained 
with  IC79 in-ice (left panels) and IT73 IceTop data (right panels). 
The in-ice data sets have median energies 
of 20~TeV (upper) and 400~TeV (lower). 
The IceTop datasets have a median energy of 
400~TeV (upper) and 2~PeV (lower). The angular binning  or smoothing angle is 20$^\circ$.}}
\end{figure}

The analysis of IceTop data confirms and complements the measurement
of the large scale anisotropy also at PeV energies\cite{Santander:2012sn}.
IceTop has an angular resolution of about 3$^\circ$
for energies above 100~TeV and zenith angles up to 40$^\circ$ 
when pure event geometry reconstruction
is performed. The angular resolution degrades to
10$^\circ$ and above for zenith angles greater than 60$^\circ$.
Only events with zenith angles less than 60$^\circ$ are
selected (1.4$\cdot$10$^8$~events/yr, $\delta >$~10$^{-4}$).
Monte Carlo studies indicate that the median primary cosmic-ray energy of 
IceTop data is 640~TeV, with 68\% of the events between
200~TeV and 2,400~TeV. Events are classified as low
(68\% of events in 100--700~TeV,  median energy of 400~TeV) 
and high energy events (68\% of events in 0.8--3.8~PeV, median energy of 2~PeV).

Analysis of IceTop data reveals a deficit that confirms what was observed 
with the in-ice muon analysis at 400~TeV (compare lower left to upper right
plot of Fig.~\ref{fig::skymaps}). The amplitude of this deficit is about 
2$\cdot$10$^{-3}$ and therefore larger than 10$^{-3}$ observed with in-ice muons. 
However, these values are in agreement if the uncertainties are considered. 
Furthermore, above 400~TeV there is indication of an increase in strength
of the anisotropy.

\section{PeV $\gamma$ search}
\label{sec::pevs}

High energy $\gamma$ rays ($\gtrsim$1~TeV) have been observed 
from different galactic sources (SNRs, pulsar wind nebulae, binary systems, 
the Galactic Center) and extra-galactic sources (starburst galaxies, 
active galactic nuclei, and objects containing supermassive black holes).
At higher energies 
($\gtrsim$100~TeV), 
extra-galactic photons are likely
to interact with the cosmic microwave background radiation 
and radiation from infrared starlight from 
galaxies, producing $e^+-e^-$ pairs. 
It is unknown whether galactic sources 
can emit $\gamma$ of energy $\gtrsim$100~TeV.
However, a guaranteed diffuse flux of $\gamma$ rays 
from interaction of cosmic rays with 
the interstellar medium and dense molecular clouds
is also expected.

IceCube can detect high energy $\gamma$ rays
by {\it vetoing} on the in-ice component, i.e.
searching for showers detected by IceTop where no in-ice
activity is observed (muon poor showers)\cite{Kolanoskiecrs}.
A PeV $\gamma$-ray shower produces about 0.1 muons above 800~GeV.
The uncertainties due to the surface detector response
and the muon rate production for photon showers
are added in quadrature and return an overall systematic uncertainty in sensitivity of 18\%.
No detectable $\gamma$-ray flux has been found by IceCube above 1.2~PeV
with one year of data.
The fraction of $\gamma$ in cosmic rays 
is estimated to be less than 1.2$\cdot$10$^{-3}$ (90\% {\it cl})
in the range 1.2--6.0~PeV and within 10$^\circ$  of the galactic plane.
IceCube is also sensitive to localized sources, where galactic accelerators or dense targets for 
extra-galactic cosmic rays might be discovered.
It is estimated that at about 1~PeV, IceCube can reveal fluxes as low as 
about 10$^{-19}$~cm$^{-2}$s$^{-1}$TeV$^{-1}$ for point sources. 

\section{Laterally separated muons}
\label{sec::lsm}

\begin{figure}[t]
\centering
\subfigure[]{\includegraphics[width=0.38\textwidth]{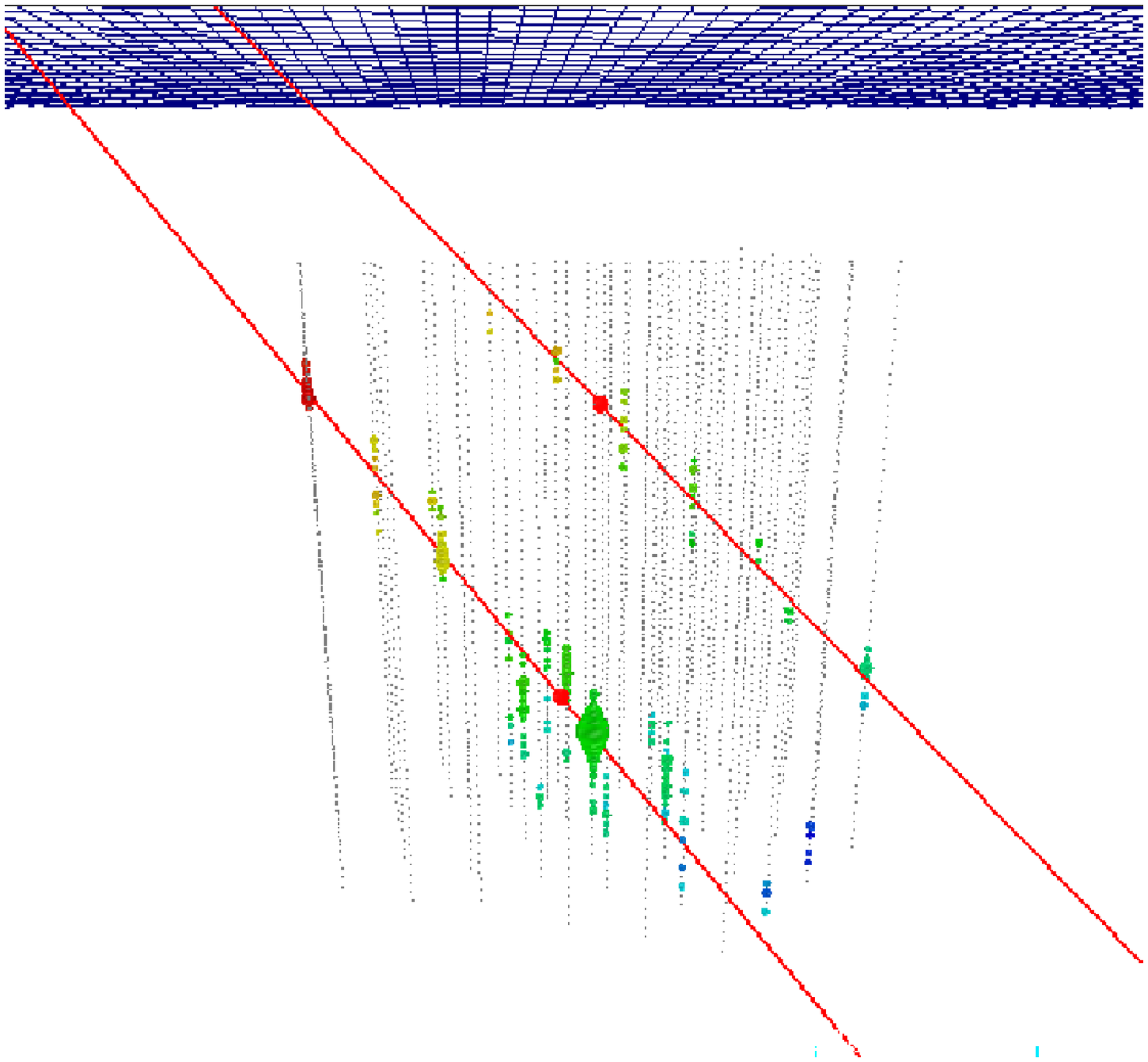}}
\subfigure[]{\includegraphics[width=0.55\textwidth]{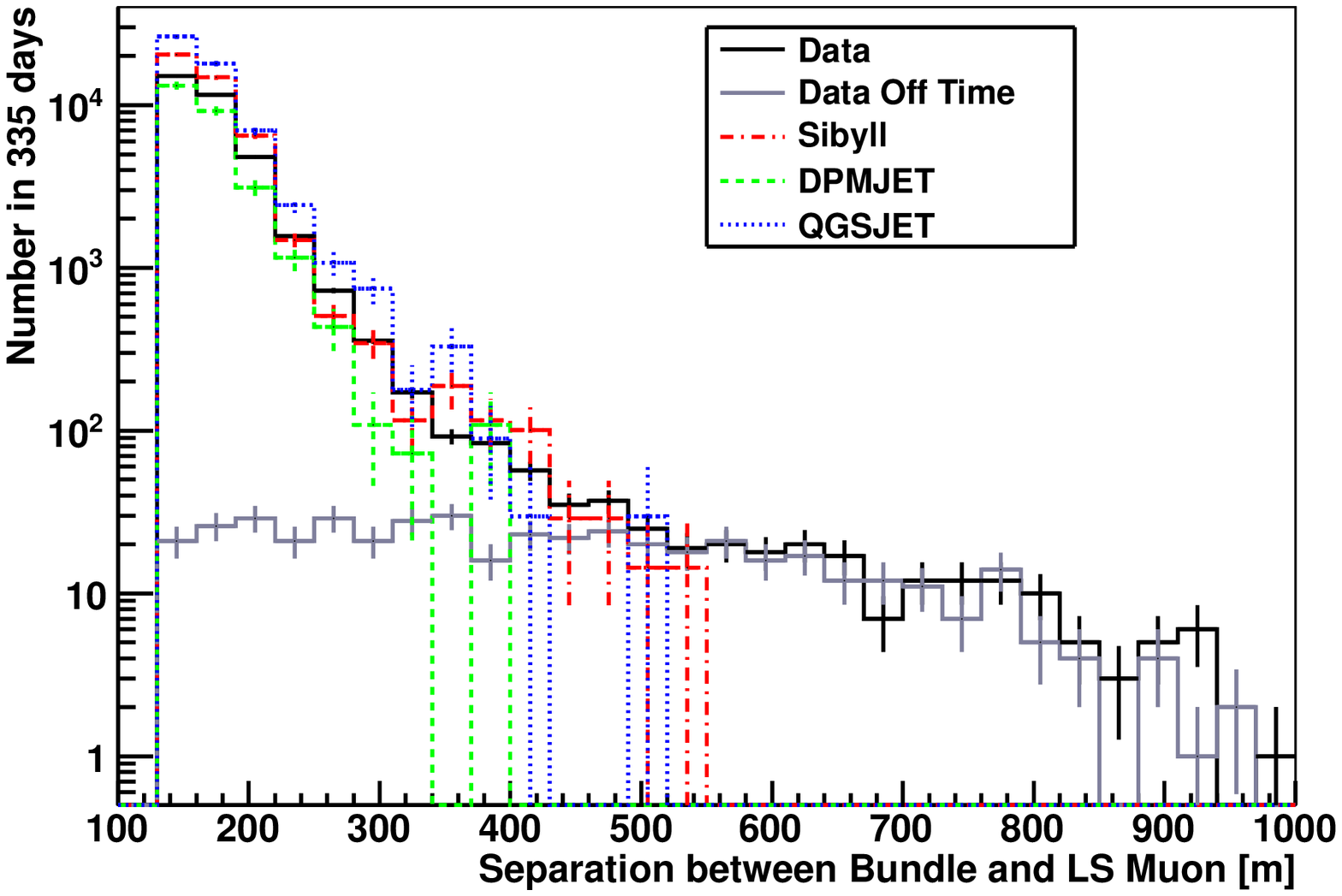}}
\caption{(a) Example of laterally separated muons (upper track) from the muon bundle
(lower track) of a cosmic-ray shower reaching the in-ice detector. The separation is 400~m,
the angular separation is 3.5$^\circ$, and the time difference is 20~ns.
(b) The separation between the laterally separated (LS) muon
and the bundle track after applying all selection criteria
for data, simulation with the Sibyll, DPMJET, and QGSJET
interaction models, as well as double showers estimated from
off-time data.}
\protect\label{fig::lsmu}
\end{figure}
High energy muons ($>$1~TeV) are produced early in air showers and
probe the initial shower development. One can distinguish between ``conventional'' muons,
which come from pion and kaon decays, and ``prompt'' muons, which
come from the decays of particles containing heavy
quarks, mostly charm. The former are expected to dominate at TeV energies,
the latter at higher but uncertain energies. Extending the earlier measurements
performed by the MACRO experiment\cite{Ambrosio:1999qu}, IceCube is capable to
resolve muons of cosmic-ray primary interactions and high-energy
secondary interactions that are laterally separated by 135~m up to 
400~m from the shower core (Fig.~\ref{fig::lsmu}), where the muon
bundle is detected\cite{Lisa:2012}.
The separation is due to high transverse momentum
of 2--15~GeV/c (corresponding to separations of 135--400~m)
transferred to the muon by its parent. Above 2~GeV/c, interactions can be described 
with perturbative quantum chromodynamics.
However, the cosmic-ray hadronic interaction models used to simulate IceCube events do not
reproduce the rates and the zenith angle distributions observed in data.

\section{Solar and Heliospheric Physics, and GRB Search}
\label{sec::moni}

\begin{figure}[t]
\centering
\includegraphics[width=10.0cm]{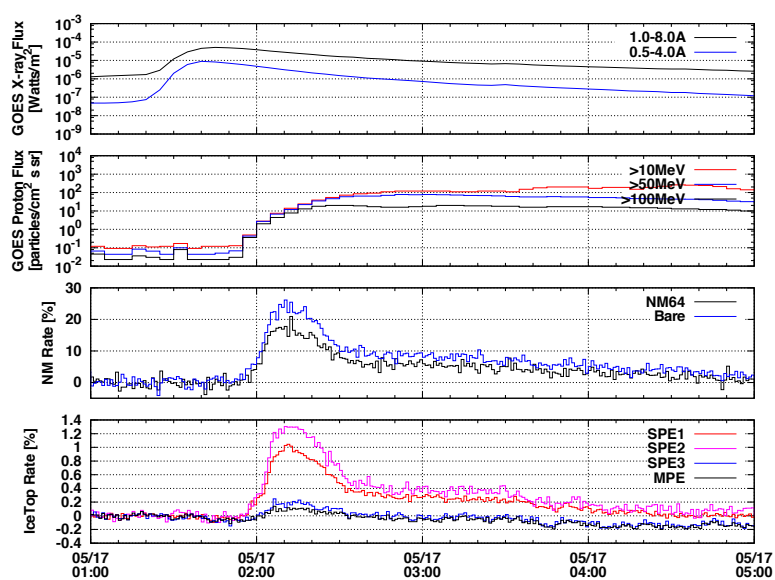}
\caption{Time profile of the count rate measured by groups of IceTop DOMs with decreasing 
thresholds (from $MPE$ to $SPE3$), neutron monitors, and the GOES satellite\protect\cite{Evenson:2012}.
\protect\label{fig::flare}}
\end{figure}
Due to the high altitude and the nearly zero geomagnetic cutoff at the South Pole, secondary
particle spectra measured with IceTop retain a significant amount of information on the spectra
of the primary particles.
IceTop has already demonstrated the novel and unique ability to
derive the energy spectrum of solar particles in the multi-GeV 
regime. The first event detected and studied by IceTop\cite{Abbasi:2008vr}
was the solar flare associated to the ground-level 
event of December 13, 2006.
IceTop DOMs can span thresholds from 1~pe to 30~pe 
corresponding to  counting rates of about 8~kHz to about 1~kHz.
By taking the differential rates at multiple thresholds,
the spectrum of detected events can be studied.
Furthermore, IceTop presents 50 times better sensitivity than conventional detectors used for solar and
heliospheric physics such as neutron monitors.
In addition to solar flares, 
the typical phenomena studied with IceTop are complex interplanetary disturbances\cite{Kuwabara:2007zz} and
``Forbush Decreases''\cite{Kuwabara:2011}. The latter are associated with strong shocks following 
coronal mass ejections that deplete cosmic rays from the region traversed by the Earth.
On May 12, 2012 a significant solar flare was detected by IceTop and is being
currently analyzed to determine the energy spectrum (Fig.~\ref{fig::flare}).

An ongoing study shows that IceTop can also reveal
GRBs through detection of an overall increase of the counting rates
observed in coincidence with the signal recorded 
on board of dedicated satellites.
A flux of $\gamma$ rays with energies greater than 10~GeV
and zenith angles up to about 20$^\circ$ can enhance the counting rate
above the steady cosmic-ray background if it is greater 
than 10$^{-5}$~erg$\cdot$cm$^{-2}$. IceTop can detect GRBs
whose emission extends up to a few 100~GeV and occurring
in a sky region not monitored by any other experiment.

\section{Conclusions}
\label{sec::conc}

The IceCube Observatory is currently taking data in the second year
after its completion in 2010. Analysis performed with data of the detector
in earlier stages of its deployment has been reviewed in this paper. 
On the one hand, events seen in coincidence by both the surface component, IceTop,
and in-ice detectors offer precious information to investigate the mass composition of cosmic rays.   
On the other hand, analysis of IceTop events allows for precise measurements
of the cosmic-ray energy spectrum with large sensitivity.
The better understanding of the systematic uncertainties along
with the exquisitely large event rate collected will allow IceCube to
significantly contribute to shed light on galactic cosmic rays
in the near future.

\section*{Acknowledgments}

I am grateful to F. Halzen for the possibility to
review IceTop results and to T. Gaisser, H. Kolanoski, and T. Stanev
for helpful discussions. A special thanks goes to
T. Feusels, L. Gerhardt, T. Kuwabara, B. Ruzybayev, and M. Santander for providing
plots.  This research is supported in part by the
U.S. National Science Foundation Grant NSF-ANT-0856253.


\end{document}